\documentstyle[11pt,aaspp4]{article}
\begin{document}
\received{ 1996 November 5}
\accepted{ 1997, September 2}
\lefthead{ Wang et al.}
\righthead{Emission Line -- Continuum Correlations}
\title{EMISSION LINE---ULTRAVIOLET TO X-RAY CONTINUUM CORRELATIONS: CONSTRAINTS ON THE 
ANISOTROPY OF THE IONIZING CONTINUUM IN ACTIVE GALACTIC NUCLEI}

\author{TING-GUI WANG, YOU-JUN LU and YOU-YUAN ZHOU}
\affil{Center for Astrophysics, University of Science and Technology of China,
Hefei, Anhui 230026, P R China}
\authoremail{tw@cfasun.cfa.ustc.edu.cn}

\begin{abstract}
Anisotropic emission of the ionizing continuum is a general prediction of the
accretion disk models. In this paper, we present the results of correlation
analysis of the UV emission line and UV to X-ray continuum properties 
for a large sample of broad emission line AGNs observed with ROSAT, IUE 
and HST. We find strong correlations between the CIV/$Ly\alpha$ ratio, 
the equivalent width of CIV, and the UV to soft X-ray spectral slope. The 
results are in good agreement with the photoionization calculation, 
suggesting that the overall ionizing continuum can well match the observed 
UV to soft X-ray spectrum. These results are consistent with the assumption of
isotropic ionizing continuum shape. Our analysis suggests a small range for 
the ``big blue bump'' cutoff energy for the objects in this sample, consistent
with the similar results of \cite{lao97,wal93} based on the continuum properties.
The mean UV-to-X-ray spectral slope is similar to the soft X-ray spectral slope.
This similarity also holds for radio-loud
and radio-quiet objects separately. This suggests that the two might be drawn from
the same distribution. The two spectral slopes are only weakly correlated. 
The UV to X-ray spectral index is correlated with absolute optical magnitude.
This result confirms the earlier suggestion that the ionizing continua are
softer for higher luminosity objects.
\end{abstract}

\section{INTRODUCTION}

It is generally accepted that the enormous energy output of Active Galactic
Nuclei is produced by the accretion of material onto putative super-massive
 black holes.  The large angular momentum present in the interstellar medium 
suggests that the accreted gas is most likely to form an accretion disk, and 
a variety of evidences for axisymmetry supports this idea.  The accretion disk 
can be of a geometrically thin, thick or intermediately slim form. The spectral
energy distribution (SED) and the intensity of the radiation from such 
disks are expected to be highly anisotropic (e.g.,\cite{cun75,lao90,mad88} 
). This effect is particularly important in the UV to X-ray band, 
which comes 
from the inner-most part of the accretion disk, where general relativistic 
effects are important. However, direct measurement of the anisotropy of 
radiation in a single object is impossible because we can only measure 
the radiation in a specific direction. An indirect measurement made using the 
very 
extended narrow emission line region suggests that the nuclear radiation is 
anisotropic (e.g., \cite{wil94}). However, the anisotropy evidenced in this 
way is expected to originate at a much larger scale (e.g., by an obscured 
torus) rather than in the nuclear region.

Broad emission lines that cover a wide range of ionization levels are thought 
to be produced by photoionization. According to the photoionization 
theory, the lines from different ionization levels respond to different parts 
of the ionizing continuum. For example, Ly$\alpha$ is produced by photons with 
energy at $\geq$13.6eV, CIV by photons with $E\geq 48$eV. Therefore the  
line spectrum provides a way of diagnosing the shape and intensity of 
the ionizing continuum. Since it is impossible to resolve the BLR 
spatially, only the total intensities of emission lines can be obtained. The 
observed line ratios and equivalent widths are dependent on some average 
shape and intensity of the ionizing continuum illuminating the BLR, which 
should be different from the observed one if the continuum emission 
is anisotropic. Therefore a comparison between the ionizing continuum 
required to 
reproduce the broad line spectrum and the observed continuum can constrain  
the role of anisotropy in the continuum emission.

For individual objects, such a comparison is very difficult to make.
First, the emission line spectrum depends on the input ionizing continuum,  
on the physical parameters and on the chemical abundances of line 
emitting gas. Since there are no
independent methods other than modeling the emission line spectrum to determine 
the physical conditions of line emitting gas, determination of the
ionizing continuum from the emission line spectrum requires accurate measurements 
of emission lines of individual elements over a wide range of ionization, which 
is usually not possible.  Second, the ionizing continuum 
at EUV energies is not directly observable because of interstellar absorptions.

However, a statistical study of this problem is feasible.  The current 
photoionization model can explain reasonably well the average observed 
strong-UV lines (\cite{net90}); hence, the average physical parameters 
can be inferred in a statistical sense. Furthermore, a great number of UV and 
soft X-ray spectra have been accumulated with IUE, HST, and ROSAT in the 
past 5 yr (e.g., \cite{wil95,zhe97,cou92,bri94}). 
The ROSAT PSPC can detect the soft X-ray spectrum down to energy of 
0.1 keV (\cite{tru83}) for low redshift AGNs. For bright high redshift 
quasars at redshift z=4, the IUE and HST spectra can access emitted 
energies of up to 40 eV in the source rest frame. 
These observations make it possible to measure the emitted continuum up to 
energies of E$\sim$10--40eV. If the intrinsic absorption is not significant 
(i.e., no strong absorption edges are seen), then the UV to X-ray spectrum can be 
roughly determined.

In this paper, we present a detailed study of the correlation between 
the broad emission line spectrum and UV to X-ray continuum properties for a 
large sample of AGNs observed 
with ROSAT/PSPC and IUE, HST. The sample and techniques of data reduction 
are presented in \S2. We present a statistical analysis in \S3. Detail 
photoionization calculations are compared with the statistical results  
in \S4, and the main conclusions are summarized in \S5.  

\section{The Sample and Data Reduction}

The heterogeneous sample consists of the 74 AGNs, the X-ray 
and UV spectra of which were analyzed for various purposes (\cite{wan96a,wan96b} 
), together with a number of AGNs observed by 
HST (\cite{lao94,lao95}). The ROSAT spectra of 39 objects in this sample
were taken from Wang, Brinkmann \& Bergeron (1996a), and 3 from Brinkmann et al. (1995). 
The X-ray spectra for the remaining 32 objects were retrieved from ROSAT 
archive 
at Max-Planck-Institut f\"ur Extraterrestrische Physik and processed in the 
manner described below.     
 
The UV spectra of 66 AGNs have been retrieved from IUE/ULDA (Uniform 
Low dispersion data archive), and were processed with IUE/SIPS. Average 
spectra have been made for those objects through multiple observations. A 
correction
was made for Galactic reddening before any line or continuum parameters 
were measured. The Galactic reddening was estimated from the neutral hydrogen 
column value given by Dickey \& Lockman (1990) with a conversion factor
 (\cite{dip94}) of
\begin{equation}
      E(B-V)=\frac{N_H^G}{5.51\times 10^{21} cm^{-2}}
\end{equation}
where $N_H^G$ is the Galactic hydrogen column density. The values of E(B-V) are listed 
in Table 1. The uncertainty in $N_H^G$ is about 
$10^{20}cm^{-2}$, corresponding to an uncertainty in $E(B-V)$ of about 0.02. 
This introduces an uncertainty of $\sim 20\%$ in the UV flux at 1350\AA. Note 
the line ratio 
CIV/Ly$\alpha$ is almost insensitive to the reddening correction for the sample 
used here. No attempt has been made to correct the intrinsic reddening, 
as it is rather uncertain. It is likely that the intrinsic reddening is small for 
our sample, since the soft X-ray fitting does not suggest significant  
absorptions above the Galactic columns (see below) for most objects.
  
The spectral indices in UV ($\alpha_{UV}$) are determined by fitting a 
power-law function ($f_\lambda \propto \lambda^{-{2+\alpha_{UV}}}$) to the 
dereddened UV spectra over several pseudo-line free windows, 1150-1180, 
1335-1365, 1450-1480, 1760-1800\AA~ (in the source rest frame). The continuum 
flux at 1350\AA~ has been determined by averaging the flux over the 
corresponding pass band, while the mean deviation is taken as uncertainty. 
The uncertainty given in this way is purely statistical and at the
$\sqrt{N-2}\sigma=2\sigma$ level (where N= width of the passband/IUE 
spectral resolution $\simeq$6 is the number of independent data points used
for taking the
average). The typical 1$\sigma$ level uncertainty for UV flux at 1350\AA~ 
is about 10-20\%. The emission line fluxes are measured by fitting the
line profiles with multiple Gaussians. For strong lines, such as CIV and 
$Ly\alpha$, usually two Gaussians are used if the line profile is symmetric, 
and three Gaussians if it is asymmetric. For the S/N 
ratios of the spectra used here, three Gaussians are a good fit to 
the data. One Gaussian  is used for fitting the weak lines, such as NV, HeII, 
SiIV+OIV], where the line centers have been fixed at the observed wavelength. 
The fitting is done locally using the IRAF package. The fitted regions are 
1180-1290, and 1480-1700\AA~ for 
$Ly\alpha$+NV and CIV+HeII, respectively. They are subject to changes when 
there is contamination in the fitting region such as geo-coronal $Ly\alpha$ 
for very low redshift AGN or when part of fitting region is shifted out of 
the spectral coverage. For the four objects, (3C351, PG1411+442, PG1351+640, 
NGC3516) with obvious associated UV line absorptions, the absorption troughs 
have been modeled with single Gaussian.   
The emission line flux is calculated by adding up all the emission line components. 
Because of the low S/N ratio of these data, the FWHM of the emission line is 
measured from the synthetic spectrum, which is constructed by putting all the 
individual 
Gaussian components together.  The FWHM measured in this way is less affected 
by the noise or the presence of weak blemishes. These results are given in 
Table~\ref{tbl-1}. As compared with the results of Wang et al. (1996b), 
the typical uncertainty due to the measurement for $CIV/Ly\alpha$ is about 
0.12. 

We have also included eight objects observed by HST in our sample. Among these, 
3C232 shows obvious absorptions in CIV and NV. The line and  
continuum parameters are simply taken from Laor et al. (1994,1995), who 
adopted a similar, but more precise model for emission lines, having 
better S/N ratio data but using a different method for estimating continuum flux. 
These lines and continuum fluxes have also been corrected for Galactic 
reddening.

\placetable{tbl-1}

The ROSAT PSPC spectra have been reduced using the EXSAS package 
(\cite{zim94}). 
The source counts were extracted from a circular region centered on the 
AGN with a radius of 3.2 arcmin. The
background was estimated from an annular source-free region.  The spectrum
was corrected for vignetting and dead-time, and regrouped to at least 20
counts per bin. The spectrum was then fitted with a single power law 
with Galactic absorption $N_{ph}(E) =A\;exp(-\sigma_E N_H) E^{-\Gamma}$, 
where $\sigma_E$ is the photoelectronic absorption cross-section
(\cite{mor83})
and $N_H$ is the absorption column density. For most objects in the sample, a reasonable 
fit can be obtained with the single power law description, and the $N_H$ 
values are consistent with the Galactic $N_H^G$ values in the corresponding
direction (\cite{dic90}). Only 7 objects (III ZW 2, I ZW 1, PKS 0405-123, 
3C 120, Mark 79, Q 1244+0240, 3C 390.3) in the sample show excessive 
absorptions 
larger than $10^{20}$~$cm^{-2}$ and at greater than 2$\sigma$ significant level.
The X-ray spectral index ($\alpha_{X}$) is related to above photon index
 $\Gamma$ by $\alpha_{X}=\Gamma-1$. The results are presented in 
Table~\ref{tbl-2}, where all error bars are quoted at 1$\sigma$ level. In 
Table~\ref{tbl-2}, we also present the absolute 
bolometric magnitudes (M$_{abs}$) of the sources, which are taken from 
Veron-Cetty \& Veron (1996).

\placetable{tbl-2}

\section{Statistical Analysis}

From the data in Table~\ref{tbl-1} and Table~\ref{tbl-2}, the UV to X-ray 
spectral slope is calculated using
\begin{equation}
\alpha_{UVX}= -0.491\; log (f_{1kev}/f_{1350}).
\end{equation}
where, $f_{1kev}$ and $f_{1350}$ are flux at 1 keV and 1350\AA~respectively.
As we have discussed in the last section, the typical uncertainty in the 
flux at 
1350\AA~ due to the measurement uncertainty plus the uncertainty in the 
Galactic reddening correction is estimated to be 40\%. The typical
 uncertainty in the 
X-ray flux at 1 keV is 10\%. Therefore a combination of these will 
introduce a typical uncertainty in $\alpha_{UVX}$ of about  0.10. 

We have also calculated the spectral slope ($\alpha_{EUV}$) between the UV 
and the soft X-ray at 0.2 keV. The uncertainty for X-ray flux at 0.2 keV 
is considerably 
larger than the corresponding error at 1 keV because the 0.2 keV flux is 
very sensitive to the absorption correction. For example, if the $N_H$ value  
varies by $10^{19} cm^{-2}$, the flux at 0.2 keV will change by 10\%. An 
uncertainty of $10^{20} cm^{-2}$ in the $N_H$~  means an uncertainty of 160\% in 
the 0.2 keV flux, or an uncertainty 0.4 in the $\alpha_{EUV}$

The distributions of $\alpha_{UVX}$ and $\alpha_{X}$ are plotted in 
Figure~\ref{fig-1}. The Kolmogorov-Smirnov test (Press et al. 1992) gives 
a probability of P=4\%. Thus the two data sets are drawn from the same 
distribution. Since the uncertainty in $\alpha_{X}$ is considerably 
larger than the uncertainty 
in $\alpha_{UVX}$,  the Kolmogorov-Smirnov test may give a false indication  because of 
variance introduced by different uncertainties in the data. To address this 
concern, we take only the objects with small uncertainties in $\alpha_{X}$. 
The Kolmogorov-Smirnov test gives a D=0.22, P=0.08 and D=0.18, P=0.32 for the subsample 
with $\sigma(\alpha_{X})<0.5$ (N=69 objects) and $\sigma_{\alpha_x}<0.3$ 
(N=55), respectively. This result suggests that the different distribution 
indicated by the Kolmogorov-Smirnov test for the whole sample is due to the objects with 
large uncertainties in $\alpha_{X}$. 

For this heterogeneous sample, the mean $<\alpha_{UVX}>=1.49\pm 0.03$ is 
in different from $<\alpha_{X}>=1.46\pm0.05$, where the error here and below is the uncertainty in the mean.  This relation $<\alpha_{X}>\simeq <\alpha_{UVX}>$ also holds when the sample is broken down into the radio-loud objects (17 radio galaxies and radio-loud quasars) where $<\alpha_{X}>=1.38\pm0.14$ and $<\alpha_{UVX}>=1.45\pm0.12$,  and to radio quiet objects where $<\alpha_{X}>=1.48\pm0.10$ and $<\alpha_{UVX}>=1.51\pm0.07$.
A similar result for $<\alpha_{OX}> \simeq <\alpha_{X}>$ has 
been noted by Brunner et al. (1992), Turner, George \& Mushotzky (1993) 
and Laor et al. (1997).
We have also checked to see whether the two spectral indices have similar medians.  
The medians are 1.53 for $\alpha_{UVX}$ and 1.39 for $\alpha_{X}$. If the uncertainty were normally distributed, 
the median would not be sensitive to the uncertainty. However, the error distribution in $\alpha_{X}$ is very likely asymmetrical, therefore the small difference in the median could be due to the larger uncertainty in the $\alpha_{X}$. Subsequent
analysis confirms this. When a subsample of objects with uncertainty in $\alpha_{X}$ less than 0.2 (41 objects) is considered, the medians for $\alpha_{UVX}$ and $\alpha_{X}$ are 1.50 and 1.48, respectively.  These 
analyses suggest that $\alpha_{X}$ and $\alpha_{UVX}$  might be drawn 
from the same distribution.

Although Laor et al. (1997) claimed that the distribution of $\alpha_{OX}$ is 
bimodal based on their sample of 23 objects and on the Figure 5b of Wang
et al. (1996b), we did not find similar result for $\alpha_{UVX}$ in our sample.   
 
Spearman rank correlation analysis has been performed among four continuum
 parameters: $\alpha_{X}$, $\alpha_{UVX}$, $\alpha_{UV}$, and the absolute 
Magnitude $M_{abs}$; and among five emission 
line parameters: Equivalent widths (EWs) and FWHM of CIV and $Ly\alpha$ and the
line ratio
CIV/$Ly\alpha$. The correlation matrix is presented in Table~\ref{tbl-3}. 
Among 36 pair of combination, we found correlation caused by
random factors with a probability less than 1\% for 24 pairs. Some correlations are obvious, such as 
between $\alpha_{UV}$ and $\alpha_{EUV}$, and between the FWHM of 
CIV and the FWHM of Ly$\alpha$. Some others have been discovered 
previously, such as the anti-correlations between $M_{abs}$ and EW(CIV), 
known as the
Baldwin effect, and between $M_{abs}$ and line ratio CIV/Ly$\alpha$, 
the positive correlations between the EWs of CIV or Ly$\alpha$ and the 
emission line width of CIV or Ly$\alpha$ (Wills et al. 1993); and the 
correlation between the line ratio CIV/Ly$\alpha$ and the line width of 
CIV or Ly$\alpha$. We will not discuss these correlations  because they 
have been extensively discussed in the literature (e.g., Wills et al. 1993; Wang et al. 1996a). 

\subsection{ Correlations among continuum properties}

The UV-to-X-ray spectral slope ($\alpha_{UVX}$) is correlated with absolute 
magnitude with a correlation coefficient $R_s=-0.50$ corresponding to 
a probability caused by a random factor of $P_r=5\times 10^{-6}$. Similarly, 
the $\alpha_{EUV}$ is also correlated with $M_{abs}$ with $R_s=-0.50$ and 
$P_r=5\times 10^{-6}$.  These correlations are similar to those
between $\alpha_{OX}$ and luminosity (e.g.,\cite{yua97}), indicating 
that the ionizing spectrum is softer when the luminosity is higher. However, 
neither the soft X-ray nor the UV spectral index is correlated with absolute 
magnitude. The $\alpha_{UVX}$ is weakly correlated with $\alpha_{X}$ (figure~\ref{fig-2}).  The 
correlation for our heterogeneous sample is substantially weaker than the one 
found in a sample of 58 bright Seyfert galaxies (\cite{wal93}). The 
correlation coefficient is only 0.44, corresponding to $P_r=2\times10^{-4}$. Fitting 
the data points to a straight line (Press et al. 1992) yields 
$\alpha_{X}=(0.61\pm0.04)+(0.60\pm0.09)\alpha_{UVX}$ with a $\chi^2=170.5$ 
for 74 data points, which is accepted at a probability of only  $Pr=6\times10^{-10}$.
We have taken into account both errors in the $\alpha_{X}$ (\cite{pre92}), with a typical error of 0.5 assigned for objects with no  error presented in Table 2,  and in the $\alpha_{UVX}$, for which a typical uncertainty of 0.1 is assigned. 
 
\subsection{Correlation between line and continuum properties}

The CIV/$Ly\alpha$ ratio is strongly correlated with $\alpha_{UVX}$ with
 $R_s=-0.58$ (see figure~\ref{fig-3}), in the sense that a flat 
UV to-X-ray spectrum 
corresponds to a larger CIV/$Ly\alpha$. This correlation is not affected 
by whether or not the four objects with associated absorption lines, PG1411+442, PG1351+640, 3C232 and 3C351 (the redshift of NGC3516 is too low to have 
$Ly\alpha$ reliably measured), are included. The UV to-X-ray 
spectra of these four absorbed objects are very steep. We have also noted that 
the other AGN with $\alpha_{UVX}>2.0$, PG0844+349 also shows an UV CIV 
absorption line in the HST FOS spectrum (\cite{cor96}).  
Furthermore, the radio-loud (RL) and radio-quiet (RQ) objects show no difference 
on the plot. 
This is consistent with the general indifference in the average 
CIV/Ly$\alpha$ ratio between RQ and RL QSOs (e.g., \cite{wil86}). 
IZW 1, Mrk478, and PG1012+008 show significantly low CIV/$Ly\alpha$ as 
compared with other 
objects with similar $\alpha_{UVX}$. Actually, their UV to X spectral slopes 
are quite normal, while their $CIV/Ly\alpha$ ratios for these three are the lowest. 
Mrk478 and I ZW 1 
are typical narrow line Seyfert 1 galaxies with strong optical FeII 
emission (\cite{bor92}); it has already noted by Wang et al.
(1996b) that strong optical FeII emitters tend to have weak CIV emission. 
However, the optical spectrum of PG1012+008 is quite normal. The low CIV/$Ly\alpha$ ratio cannot be due to measurement error. The CIV line in the spectrum of 
PG1012+008 as processed with the method of optimal extraction also appears 
extremely weak on the plot of Lanzetta, Turnshek, \& Sandoval (1993).
A correlation between the CIV/$Ly\alpha$ ratio and the soft X-ray spectral 
slope is also found, with $R_s=-0.43$, corresponding to $P_r=1\times 10^{-4}$. 

The EW of CIV is much better correlated with $\alpha_{UVX}$ 
($R_s=-0.59$, $P_r=3\times 10^{-8}$) than the $Ly\alpha$ EW ($R_s=-0.43$, 
$P_r=2\times 10^{-4}$) (figure~\ref{fig-4}) is.
The CIV EW appears to be also correlated with 
$\alpha_{X}$ ($R_s=-0.46$, $P_r=5\times 10^{-5}$) and $Ly\alpha$ EW appears to be correlated 
with $\alpha_{UV}$ ($R_s=-0.38$, $P_r=0.001$). 

The correlations of line parameters with $\alpha_{EUV}$ are similar to 
those with $\alpha_{UVX}$ (see Table 3). The slightly lower correlation 
coefficients are probably due to the larger uncertainties in the 0.2keV flux 
caused by the uncertainty in the absorption correction.

\section {DISCUSSION}

We have found strong correlations between the CIV/$Ly\alpha$ ratio, 
the CIV EW, and the observed UV to X-ray continuum 
shape for a large sample of AGNs observed by ROSAT/PSPC and IUE, HST. 
These correlations have also been noted for a small sample of AGNs 
(\cite{sch92,gre96}).
Similar correlations between the OVI/$Ly\alpha$ ratio, OVI EW, and 
the UV to X-ray spectral slope have been reported by Zheng et al. (1995)
for a small sample of moderate-redshift AGN observed by IUE, the Hopkins Ultraviolet Telescope (HUT) and 
HST. In this section we will discuss the implications of these results.  

The predicted emission-line spectrum is sensitive to the exact shape of the 
``blue bump'' in the UV to soft X-ray, since it is the photons at these 
energies 
that govern the ionization level and the excitation of the most commonly observed 
emission lines (e.g., \cite{bin89}). In addition, Ferland et al. (1992)
showed that the spectral shape at 0.1-1mm can also alter the strength of 
collisional-excited line CIV.
That the observed CIV/$Ly\alpha$ decreases with increasing steepness of the ionizing
 continuum shape in UV to X-ray is in qualitative agreement with expectation of
photoionization models. One obvious suggestion is that the observed 
continuum shape is related to average ionizing continuum of BLR.  
In the worst case, i.e., we assume that all the dispersions in the relation 
are due to intrinsic continuum anisotropy (see Fig.~\ref{fig-4}), and thus 
the maximum anisotropy produces a scatter in $\alpha_{UVX}\simeq 1$. 
However, the actual situation is much better than this (see below).
 
In order to see how the observed data match the photoionization prediction, 
we have made a series of photoionization calculations using Cloudy 84.12 
(\cite{fer94}). We have assumed solar abundances, because CIV is the main 
coolant in the ionized hydrogen (HII) zone and because CIV emission is mainly 
determined by the heating rate at that zone; therefore the CIV/$Ly\alpha$ ratio is 
not very sensitive to the assumed chemical abundances (\cite{dn79}). 
The hydrogen column density is assumed to be $10^{23}$~cm$^{-2}$. Other 
values should produce similar CIV/Ly$\alpha$, provided the gas is not 
too optically thin. The ionization parameter (the ratio of the ionizing 
photon density and particle density) and particle density range from $10^{-2.5}$ to 0.1 and from $10^9$ 
to $10^{10}$~cm$^{-3}$, respectively. The input ionizing continuum is 
approximated by a power law between the far UV and 1 keV with varying spectral 
slope, similar to the composite EUV spectrum found by Zheng et al. 
(1997) for a sample of quasars observed by HST,   
 but otherwise similar to the ``mean AGN spectrum'' of Mathews \& 
Ferland (1987). In addition, the geometry of broad emission line cloud
 distribution is assumed to be spherical, so the line intensity is the average 
of the emission from the illuminated and the back surfaces of the clouds
. If the distribution of broad line clouds were not  
symmetrical geometry, then the line emission would be anisotropic, and the 
observed line ratios would depend on the inclination. For optically thick clouds, 
the Ly$\alpha$ photons are more 
forward beamed than CIV photons are (Ferland et al. 1992). Consequently, the
line ratio CIV/$Ly\alpha$ is larger when one looks at the illuminated face, 
and smaller at the back face. Nevertheless, as long as some rotational symmetry is kept for the geometry and as long as there are no preferred absorptions on either the far 
or the near side of the clouds, the line ratios are similar  for the 
spherical symmetric geometry (e.g., \cite{kal86}). The calculated CIV/Ly$\alpha$~ are plotted against $\alpha_{UVX}$ in
figure~\ref{fig-3}. 
 
Obviously, the observed CIV/Ly$\alpha$ versus $\alpha_{UVX}$ can be reproduced 
with a relatively narrow range of physical parameters. The result
shows that CIV/Ly$\alpha$ ratio is much more sensitive to the physical 
conditions when the UV to X spectrum is harder. This is in good agreement 
with interpretation of the scatter in terms of different physical parameters. 
For $n_H=10^{10}cm^{-3}$, the range in U is about a factor of 20. This 
range is consistent with the results of reverberation mapping of bright AGNs 
for which the BLR size deviates from $R\propto L^{1/2}$ by a factor of sometimes greater than 5 (e.g., \cite{pet93,kas97}). Therefore, this interpretation is sensible. And if 
the typical measurement error for CIV/Ly$\alpha\sim 0.12$ is 
taken into account, the incident ionizing continuum on BLR must be very close 
to what we see. The UV to X-ray spectrum is likely not a power law, but a 
combination of ``blue bump'' and a hard power-law component. A different 
cutoff energy of the ``bump'' will further introduce scatter in the 
correlation (see below). A more detailed analysis requires higher quality data. 

The correlation between the $Ly\alpha$ EW and the $\alpha_{UV}$ is a 
natural prediction if the observed UV spectrum extends to the Lyman limit 
and the BLR sees the same continuum as we do. However, the correlation 
between the $Ly\alpha$ EW and $\alpha_{UVX}$ requires that continuum 
slope in the Lyman continuum range also be related to the $\alpha_{UVX}$ since 
$Ly\alpha$ is produced mainly by recombination process. 
Since $\alpha_{UVX}$ is not correlated with $\alpha_{UV}$ (see \S3), this 
requires that the UV continuum extend to the Lyman limit to produce the former 
correlation, but not to too high energy, or it would destroy the latter 
. The fact that $\alpha_{UV}$ is not correlated with $CIV$ EW also 
suggests that the UV spectrum does not extend too far to the $Ly\alpha$ limit 
continuum.

For the description of the ``big blue bump'' as $f_\nu\propto\nu^{-\alpha_{UV}} 
e^{-h\nu/E_{cut}}$, it is shown that the range in the cutoff energy, 
$E_{cut}$, is small (\cite{wal93,lao97}). 
With this range of $E_{cut}$, the CIV ionizing 
photon numbers do not change much; however, the heating rate changes 
significantly if the $\alpha_{UVX}$ is soft. As a result, the CIV/Ly$\alpha$ 
ratio is sensitive to the cutoff energy of the ``big blue bump'' (cf., \cite{bin89}). 
Grid photoionization calculations are made for $n_H=10^{10}$~cm$^{-3}$, 
$N_c=10^{23}$~cm$^{-2}$, and solar chemical abundance, with results   
in agreement with the above qualitative analysis (see figure~\ref{fig-5}). For example, for $\alpha_{UVX} \sim1.5-2.0$, the difference in CIV/$Ly\alpha$ 
ratio is 0.28 if $E_{cut}$ vary from 30eV to 60 eV, 
which is similar to the value of the observed scatter in CIV/$Ly\alpha$ versus 
$\alpha_{UVX}$ 
correlation. A larger cut-off energy 
range would produce a scatter larger than the one observed. This analysis 
provides an independent evidence for a small range of cut-off energy.  


Francis (1993) and Netzer, Laor \& Gondhalekhar (1992) found that the distributions 
of CIV and $Ly\alpha$ EWs are narrower than those expected for an ionizing 
continuum source from a randomly inclined accretion disk. Here, we show that 
the observed $\alpha_{UVX}$ distribution can contribute a similar size 
to the scatter of ${Ly\alpha}$ and CIV EWs. Taking away the $\alpha_{UVX}$ factor, the distribution for 
$Ly\alpha$ EW is narrowed down significantly. If one accepts the 
rough isotropy of UV to X-ray spectrum we argued above, the random thin 
disk distribution of ionizing continuum strength is even more 
problematic. 

\section{Conclusion}

We have presented the results of correlation analysis of UV and soft X-ray spectra for a large sample of AGN. The main results are summarized as 
follows:

\begin{enumerate}
\item The UV-to-X-ray spectral indices are strongly correlated with line ratios 
CIV/$Ly\alpha$. This correlation can be modeled with photoionization models 
that assume ionizing continua with a range of UV to X-ray spectral slopes, and the scatter can be interpreted as due to the uncertainties in the physical conditions of the BLR. We 
suggest that the average ionizing spectrum striking the BLR is similar to the observed one. If the ionizing continuum consists of a power law and the ``big blue bump'' 
components, the range of bump cutoff energy must be small in order to be
consistent with the correlation.

\item The UV to X-ray spectral indices are significantly correlated with the 
EWs of CIV and Ly$\alpha$. These correlations are also consistent with the 
the intepretation of isotropic ionizing continuum shape. As a consequence of this correlation, the constraints put by Netzer et al. (1992) and Francis (1993) on the anisotropy of the continuum would be even stronger.

\item The UV to X-ray spectral index is correlated with absolute optical magnitude. This result confirms the earlier suggestion that the ionizing 
continua are softer for objects with higher luminosities. 

\item The mean UV to X-ray spectral slope is similar to the soft X-ray 
spectral slope. This similarity also holds for radio-loud and radio-quiet 
objects separately. This suggests that the two may be drawn from the same 
distribution. The two spectral slopes are only weakly correlated. 

\end{enumerate}

\acknowledgments
We thank the anonymous referee for many useful suggestions that significantly 
improved the presentation of this paper. This work is partly supported by the
Chinese Natural Science Foundation and the Panden Project.

\clearpage
\figcaption[fig1.eps]{The distributions of UV-to-X-ray spectral slope and the soft X-ray spectral slope. The solid line is for the $\alpha_{X}$, whereas 
the dashed line for $\alpha_{UVX}$. Typical sizes of error bar are 0.2 and  
0.5 for $\alpha_{EUV}$ and $\alpha_{X}$, respectively. \label{fig-1}}

\figcaption[fig2.eps]{The ROSAT PSPC spectral indices $\alpha_{X}$ versus 
the UV-to-X-ray spectral index ($\alpha_{UVX}$). For the data from ROSAT 
All Sky Survey, the error bars for X-ray spectral index are not shown. 
The dashed line in the figure is $\alpha_{X}$=$\alpha_{UVX}$. \label{fig-2} }

\figcaption[fig3.eps]{The correlation of line ratio  CIV/$Ly\alpha$ versus UV-to-X-ray spectral index. Filled squares are for radio quiet objects, 
open circles are for the radio-loud objects. The theoretical predictions 
are shown as curves for hydrogen particle densities $10^{10}$~cm$^{-3}$
(solid line) and $10^{9}$~cm$^{-3}$ (dashed-line). For each curve, the ionization parameters are labeled. Typical error bar is shown in the 
upper-right corner.   \label{fig-3}}

\figcaption[fig4.eps]{Correlations of line equivalent width with $\alpha_{UVX}$. The $\alpha_{UVX}$ versus EW CIV correlation is stronger than the $\alpha_{UVX}$
versus EW Ly$\alpha$ correlation is. \label{fig-4}}

\figcaption[fig5.eps]{Illustration of the impact of cutoff energy of the ``big blue bump'' 
on the line ratio CIV/Ly$\alpha$ for various UV-to-X-ray spectral indices. The models adopted here use ionization parameter, U=0.01, column density $N_H=10^{23}cm^{-2}$ and particle density $n_H=10^{10}cm^{-3}$. \label{fig-5}}
 
\begin{table}
\dummytable\label{tbl-1}
\end{table}
\begin{table}
\dummytable\label{tbl-2}
\end{table}
\begin{table}
\dummytable\label{tbl-3}
\end{table}
\end{document}